\documentclass[a4paper,11pt,reqno]{amsart}

\usepackage[a4paper]{geometry}
\usepackage{amssymb,amsmath}

\usepackage{mathtools}
\usepackage{enumerate}
\usepackage{bookmark}
\usepackage{hyperref}
\usepackage[initials,backrefs]{amsrefs}

\numberwithin{equation}{section} 

\theoremstyle{plain}
\newtheorem{theorem}{Theorem}

\theoremstyle{definition}
\newtheorem{definition}{Definition}
\newtheorem*{assi*}{(I) Infinite range interaction}
\newtheorem*{assp*}{(P) Singular distributions}
\newtheorem*{dskn*}{$\dskn$}
\newtheorem*{dsknn*}{$\dsknn$}

\theoremstyle{remark}

\newcommand{\prob}[1]{\DP\left\{#1\right\}}
\newcommand{\esm}[1]{\mathbb{E}\left[\,#1\,\right]}

\newcommand{\Bone}{\mathbf{1}}

\newcommand{\BC}{\mathbf{C}}

\newcommand{\BG}{\mathbf{G}}
\newcommand{\BH}{\mathbf{H}}

\newcommand{\BK}{\mathbf{K}}

\newcommand{\BP}{\mathbf{P}}
\newcommand{\BU}{\mathbf{U}}
\newcommand{\BV}{\mathbf{V}}

\newcommand{\BX}{\mathbf{X}}

\newcommand{\CB}{\mathcal{B}}
\newcommand{\CJ}{\mathcal{J}}

\newcommand{\DP}{\mathbb{P}}
\newcommand{\DR}{\mathbb{R}}
\newcommand{\DZ}{\mathbb{Z}}

\newcommand{\BDelta}{\mathbf{\Delta}}
\newcommand{\BPsi}{\mathbf{\Psi}}
\newcommand{\Bx}{\mathbf{x}}
\newcommand{\By}{\mathbf{y}}

\newcommand{\Bu}{\mathbf{u}}
\newcommand{\Bv}{\mathbf{v}}
\newcommand{\FB}{\mathfrak{B}}

\DeclareMathOperator{\card}{card}

\DeclareMathOperator{\dist}{dist}
\DeclareMathOperator{\supp}{supp}

\newcommand{\ee}{\mathrm{e}}

\newcommand{\condI}{\mathbf{(I)}}
\newcommand{\condP}{\mathbf{(P)}}

\begin{document}

\title[Multi-particle localization with infinite range interaction]{On complete localization for the one-dimensional multi-particle Anderson-Bernoulli model with infinite range interaction} 

\author[T.~Ekanga]{Tr\'esor EKANGA$^{\ast}$}

\address{$^{\ast}$%
Institut de Math\'ematiques de Jussieu,
Universit\'e Paris Diderot,
Batiment Sophie Germain,
13 rue Albert Einstein,
75013 Paris,
France}
\email{ekanga@math.jussieu.fr}
\date{\today}
\begin{abstract}
We consider the multi-particle Anderson model on the lattice with infinite range but sub-exponentially decaying interaction  and show the Anderson localization consisting of the spectral exponential and the strong dynamical localization. In particular, the dynamical localization is proved in the Hilbert-Schmidt norm. The results concern very singular probability distributions such as the Bernoulli's measures.
\end{abstract}

\maketitle

\section{Introduction}
Multi-particle quantum disordered systems with infinite range inter-particle interaction were recently analyzed on the one hand by Fauser and Warzel \cite{FW15} in the continuous space and under either the low energy or the weak interaction regime with the help of an adapted version of the fractional moment method. On the other hand, Chulaevsky himself \cite{C11}, proved localization first, for the multi-particle system with long-range interaction on the lattice. Second, in the continuum under the low energy regime \cite{C14}.

All the previous works on localization for many body interacting quantum particles systems, \cites{AW09,CS09,E11,E12,E17b}, the interaction potential was assumed of a finite range. In the recent papers by Chulaevsky \cites{C11,C14}, the author used a quite modified version of the variable energy multi-scale analysis for multi-particle systems.

In the analysis by Fauser and Warzel \cite{FW15} as well as the one by Chulaesvky \cite{C11}, the infinite range inter-particle interaction was assumed to be sub-exponentially decaying fast at infinity.

In the present work, we treat the weak interaction regime and prove localization with the help of the multi-particle multi-scale analysis developed in our earlier work \cite{E17b}. In both the two localization regimes, the common probability distribution of the i.i.d. random external potential in the Anderson model is allowed to be singular including Bernoulli's measures. This is an important advantage of the multi-scale analysis in contrast to the fractional moment method in the general theory of the mathematics of Anderson localization. The fractional moment method usually requires regular distributions such as absolutely continuous with a bounded density. The first mathematical proofs of the Anderson localization with singular Bernoulli distributions were obtained for single-particle models on the lattice by Carmona et al. \cite{CKM87} and in the continuum by Damanik et a. \cite{DSS02}. 

Our main results for the weak interaction regime are Theorem \ref{thm:weak.interaction.exp.loc} and \ref{thm:weak.interaction.dynamical.loc}.
The proofs use the Wegner type-bounds for Bernoulli distributions proved in the work \cite{E17a} combined with the multi-scale analyzes of \cite{E17b} for the weak interaction regime.

\section{The model, hypotheses and the main results}

\subsection{The model}
We define the two following norms on $\DR^D$ for arbitrary $D\geq 1$: $|\Bx|=\max_{i=1,\ldots,D}|x_i|$ and $|\Bx|_1=|x_1|+\cdots+|x_D|$ and we will use the inclusion $\DZ^D\subset \DR^D$. The first norm will be used  to define the cubes  in $\DZ^d$ and the second in the definition of the lattice Laplacian.  We consider a system of $N$-particles where $N\geq 2$ is finite and fixed. Let $d\geq 1$ and  $1\leq n\leq N$. We analyze random Hamiltonian $\BH^{(n)}_{h}(\omega)$  of the form
\begin{equation}\label{eq:hamiltonian}
\BH^{(n)}_{h}(\omega)=-\BDelta+\sum_{j=1}^nV(x_j,\omega)+h\BU=-\BDelta+\BV(\Bx,\omega)+h\BU,
\end{equation}
acting on $\ell^2((\DZ^{d})^n)\cong \ell^2(\DZ^{nd})$ with $h\in\DR$ and $\Bx\in(\DZ^d)^n$. Above, $\BDelta$ is the $nd$-dimensional lattice nearest-neighbor Laplacian:
\begin{equation}           \label{eq:def.Delta}
(\BDelta\BPsi)(\Bx)
=\sum_{\substack{\By\in\DZ^{nd}\\|\By-\Bx|_1=1}}\left(\BPsi(\By)-\BPsi(\Bx)\right)
=\sum_{\substack{\By\in\DZ^{nd}\\|\By-\Bx|_1=1}}\BPsi(\By)-2dn\BPsi(\Bx),
\end{equation}
for $\BPsi\in\ell^2(\DZ^{nd})$ and $\Bx\in\DZ^{nd}$. $V\colon\DZ^d\times\Omega\to\DR$ is a random field relative to a probability space $(\Omega,\FB,\DP)$ and $\BU\colon(\DZ^d)^n\to\DR$ is the potential of inter-particle interaction. $\BV$ and $\BU$ act on $\ell^2(\DZ^{nd})$ as multiplication operators by functions $\BV(\Bx,\omega)$ and $\BU(\Bx)$ respectively.

\subsection{The hypotheses}

\begin{assi*}

Fix any $n=1,\ldots,N$. The potential of inter-particle interaction $\mathbf{U}$ is bounded and of the form
\[
\BU(\Bx)=\sum_{1\leq i<j\leq n} U(x_i,x_j),\quad \Bx=(x_1,\ldots,x_n),
\]
where  $U:\DZ^2\rightarrow\DR$ is a function such that there exist positive constants $c,C$ and some $0<\tau<1$ such that: 

\begin{equation}\label{eq:finite.range.k}
\forall x,y\in\DZ, \qquad U(x,y)\leq C\ee^{-c|x-y|^{\tau}}.
\end{equation}

\end{assi*}
Let us make some comments on this assumption. In the case of the interaction of infinite range, the assumption $\condI$ is combined  with the second resolvent identity in order to bound some probability of the multi-scale analysis. 

Set $\Omega=\DR^{\DZ^d}$ and $\FB=\bigotimes_{\DZ^d}\mathcal{B}(\DR)$ where $\mathcal{B}(\DR)$ is the Borel sigma-algebra on $\DR$. Let $\mu$ be a probability measure on $\DR$ and define $\DP=\bigotimes_{\DZ^d}\mu$ on $\Omega$.

The external random potential $V\colon\DZ^d\times\Omega\to\DR$ is an i.i.d. random field  
relative to  $(\Omega,\FB,\DP)$ and is defined by $V(x,\omega)=\omega_x$ for $\omega=(\omega_i)_{i\in\DZ^d}$.
The common probability distribution function, $F_V$,  of the i.i.d. random variables $V(x,\cdot)$, $x\in\DZ^d$ associated to the measure $\mu$ is defined by 
\[
F_V: t \mapsto \prob{V(0,\omega)\leq t }.
\]
\begin{assp*}
The random potential $V:\DZ^d\times\Omega\rightarrow \DR$ is almost surely bounded, i.e., there exists $M\in(0,\infty)$ such that $0\in\supp \mu \subset[-M,M]$ where $\mu$ is the common probability distribution measure of the random field $\{V(x,\omega)\}_{x\in\DZ^d}$. Further, $\supp \mu$ is not concentrated in a single point and $\int|x|^{\eta}d\mu(x)<+\infty$ for some $\eta>0$.
\end{assp*}

\subsection{The results}

\begin{theorem}\label{thm:weak.interaction.exp.loc}
Let $d=1$. Under assumptions $\condI$ and $\condP$,
 there exists $h^*>0$ such that for any $h\in(-h^*, h^*)$ the Hamiltonian $\BH^{(N)}_h$, with interaction of amplitude $|h|$, exhibits complete Anderson localization, \emph{i.e.}, with $\DP$-probability one,
the spectrum of $\BH^{(N)}_h$  is pure point, and all the eigenfunctions $\BPsi_i(\Bx,\omega)$ are exponentially decaying at infinity: 
\[
|\BPsi_i(\Bx,\omega)|\leq C_i\ee^{-c|\Bx|}, 
\]
for some positive constants $c$ and $C_i$. 
\end{theorem}

Denote by  $\CB_1$ the set  of bounded measurable functions $f:\DR\rightarrow\DR$ such that $\|f\|_{\infty}\leq 1$. We now give our result on strong Hilbert-Schmidt dynamical localization.

\begin{theorem}\label{thm:weak.interaction.dynamical.loc}  
Let $d=1$. Under assumptions $\condI$ and $\condP$, there exists $h^*>0$ and $s^*>0$ such that for any $h\in(-h^*, h^*)$  any bounded Borel function $f:\DR\rightarrow\DR$, any bounded region $\BK\subset\DZ^{Nd}$ and any $s\in(-s^*,s^*)$ we have: 
\begin{equation}\label{eq:weak.interaction.dynamical.loc} 
\esm{\sup_{f\in\CB_1}\Bigl\| |\BX|^{\frac{s}{2}}f(\BH^{(N)}(\omega))\BP_{I}(\BH^{(N)}_h(\omega))\Bone_{\BK}\Bigr\|_{HS}^2}<\infty,
\end{equation}
where $(|\BX|\BPsi)(\Bx):=|\Bx|\BPsi(\Bx)$, $\BP_{I}(\BH^{(N)}(\omega))$ is the spectral projection of $\BH^{(N)}(\omega)$ onto the interval $I$.
\end{theorem}

\section{The general strategy of the proofs}

\subsection{Geometry}
For $\Bu=(u_1,\ldots,u_n)\in\DZ^{nd}$, we denote by $\BC^{(n)}_L(\Bu)$ the $n$-particle cube, i.e, 
\[
\BC^{(n)}_L(\Bu)=\left\{\Bx\in\DZ^{nd}:|\Bx-\Bu|\leq L\right\},
\]
and given $\{L_i: i=1,\ldots,n\}$, we define the rectangle
\begin{equation}          \label{eq:cube}
\BC^{(n)}(\Bu)=\prod_{i=1}^n C^{(1)}_{L_i}(u_i),
\end{equation}
where $C^{(1)}_{L_i}(u_i)$ are cubes of side length $L_i$ center at points $u_i$.
We  define the \emph{internal boundary} of the domain $\BC^{(n)}(\Bu)$ by
\begin{equation}\label{eq:int.boundary}
\partial^-\BC^{(n)}(\Bu)=\left\{\Bv\in\DZ^{nd}:\dist\left(\Bv,\DZ^{nd}\setminus \BC^{(n)}(\Bu)\right)=1\right\},       
\end{equation}
and its \emph{external boundary} by
\begin{equation}\label{eq:ext.boundary}
\partial^+\BC^{(n)}(\Bu)=\left\{\Bv\in\DZ^{nd}\setminus\BC^{(n)}(\Bu):\dist\left(\Bv, \BC^{(n)}(\Bu)\right)=1\right\}.
\end{equation}
The cardinality of the cube $\BC^{(n)}_L(\Bu)$ is $|\BC_L^{(n)}(\Bu)| := \card\BC_L^{(n)}(\Bu)=(2L+1)^{nd}$.
We define the restriction of the Hamiltonian $\BH_h^{(n)}$ to  $\BC^{(n)}(\Bu)$ by
\begin{align*}
&\BH_{\BC^{(n)}(\Bu),h}^{(n)}=\BH^{(n)}_h\big\vert_{\BC^{(n)}(\Bu)}\\
&\text{with simple boundary conditions on }\partial^+\BC^{(n)}(\Bu),
\end{align*}
i.e., $\BH^{(n)}_{\BC^{(n)}(\Bu),h}(\Bx,\By)=\BH^{(n)}_h(\Bx,\By)$ whenever $\Bx,\By\in\BC^{(n)}(\Bu)$ and $\BH^{(n)}_{\BC^{(n)}(\Bu),h}(\Bx,\By)=0$ otherwise.
We denote the spectrum of $\BH_{\BC^{(n)}(\Bu),h}^{(n)}$  by
$\sigma\bigl(\BH_{\BC^{(n)}(\Bu)}^{(n),h}\bigr)$ and its resolvent by
\begin{equation}\label{eq:def.resolvent}
\BG_{\BC^{(n)}(\Bu),h}(E):=\Bigl(\BH_{\BC^{(n)}(\Bu),h}^{(n)}-E\Bigr)^{-1},\quad E\in\DR\setminus\sigma\Bigl(\BH_{\BC^{(n)}(\Bu),h}^{(n)}\Bigr).
\end{equation}
The matrix elements $\BG_{\BC^{(n)}(\Bu),h}(\Bx,\By;E)$ are usually called the
\emph{Green functions} of the operator $\BH_{\BC^{(n)}(\Bu),h}^{(n)}$.

\begin{definition}
Let $m>0$ and $E\in\DR$ be given.  A cube $\BC_L^{(n)}(\Bu)\subset\DZ^{nd}$, $1\leq n\leq N$ will be  called $(E,m,h)$-\emph{nonsingular} ($(E,m,h)$-NS) if $E\notin\sigma(\BH^{(n)}_{\BC^{(n)}_{L}(\Bu),h})$ and
\begin{equation}\label{eq:singular}
\max_{\Bv\in\partial^-\BC_L^{(n)}(\Bu)}\left|\BG_{\BC_L^{(n)}(\Bu),h}(\Bu,\Bv;E)\right|\leq\ee^{-\gamma(m,L,n)L},
\end{equation}
where 
\begin{equation}\label{eq:gamma}
\gamma(m,L,n)=m(1+L^{-1/8})^{N-n+1}.
\end{equation}
Otherwise it will be called $(E,m,h)$-\emph{singular} ($(E,m,h)$-S).
\end{definition}

We will also make use of the following notion.

\begin{definition}\label{def:separability}
A cube $\BC^{(n)}_L(\Bx)$ is  $\CJ$-separable from $\BC^{(n)}_L(\By)$ if there exists a nonempty subset $\CJ\subset\{1,\cdots,n\}$ such that
\[
\left(\bigcup_{j\in \CJ}C^{(1)}_{L}(x_j)\right)\cap
\left(\bigcup_{j\notin \CJ}C_L^{(1)}(x_j)\cup \bigcup_{j=1}^n C_L^{(1)}(y_j)\right)=\emptyset.
\]
A pair $(\BC^{(n)}_L(\Bx),\BC^{(n)}_L(\By))$ is separable if $|\Bx-\By|>7NL$ and if one of the cube is $\CJ$-separable from the other.
\end{definition}

\subsection{The multi-scale analysis}

Rhe multi-scale analysis bounds is summarized in the following result:
\begin{theorem}\label{thm.MSA}
For any pair of separable cubes $\BC_{L_k}^{(n)}(\Bu)$ and $\BC_{L_k}^{(n)}(\Bv)$
\begin{equation}\label{eq:property.DS.k.n.N}
\DP\left\{\exists\,E\in I:\BC_{L_k}^{(n)}(\Bu),\ \BC_{L_k}^{(n)}(\Bv) \text{ are } (E,m)\text{-S}\right\}
\leq L_k^{-2p\,4^{N-n}},
\end{equation}
where $m>0$, $p>6Nd$, and $I\subset \DR$ is a compact interval.
\end{theorem}
For the proof, we refer to the papers \cites{E12,E17b}.

\subsection{Proof of the results}
The proofs of the localization results given the multi-scale analysis bounds Theorem \ref{thm.MSA}, are done is the same way as in the works \cites{E12,E17b}

\end{document}